\PassOptionsToPackage{unicode}{hyperref}
\PassOptionsToPackage{hyphens}{url}
\PassOptionsToPackage{dvipsnames,svgnames,x11names}{xcolor}
\documentclass[
  10pt,
  twocolumn]{paper}
\usepackage{amsmath,amssymb}
\usepackage{iftex}
\ifPDFTeX
  \usepackage[T1]{fontenc}
  \usepackage[utf8]{inputenc}
  \usepackage{textcomp} 
\else 
  \usepackage{unicode-math} 
  \defaultfontfeatures{Scale=MatchLowercase}
  \defaultfontfeatures[\rmfamily]{Ligatures=TeX,Scale=1}
\fi
\usepackage{lmodern}
\ifPDFTeX\else
\fi
\IfFileExists{upquote.sty}{\usepackage{upquote}}{}
\IfFileExists{microtype.sty}{
  \usepackage[]{microtype}
  \UseMicrotypeSet[protrusion]{basicmath} 
}{}
\makeatletter
\@ifundefined{KOMAClassName}{
  \IfFileExists{parskip.sty}{%
    \usepackage{parskip}
  }{
    \setlength{\parindent}{0pt}
    \setlength{\parskip}{6pt plus 2pt minus 1pt}}
}{
  \KOMAoptions{parskip=half}}
\makeatother
\usepackage{xcolor}
\usepackage[margin=0.98in]{geometry}
\usepackage{color}
\usepackage{fancyvrb}

\DefineVerbatimEnvironment{Highlighting}{Verbatim}{commandchars=\\\{\}}
\newenvironment{Shaded}{}{}

\newcommand{\DataTypeTok}[1]{#1}

\newcommand{\FunctionTok}[1]{#1}

\newcommand{\OtherTok}[1]{\textcolor[rgb]{1.00,0.25,0.00}{#1}}

\newcommand{\StringTok}[1]{\textcolor[rgb]{0.00,0.50,0.50}{#1}}

\usepackage{longtable,booktabs,array}
\usepackage{calc} 
\usepackage{etoolbox}
\makeatletter
\patchcmd\longtable{\par}{\if@noskipsec\mbox{}\fi\par}{}{}
\makeatother
\IfFileExists{footnotehyper.sty}{\usepackage{footnotehyper}}{\usepackage{footnote}}
\makesavenoteenv{longtable}
\usepackage{graphicx}
\makeatletter
\def\maxwidth{\ifdim\Gin@nat@width>\linewidth\linewidth\else\Gin@nat@width\fi}
\def\maxheight{\ifdim\Gin@nat@height>\textheight\textheight\else\Gin@nat@height\fi}
\makeatother
\setkeys{Gin}{width=\maxwidth,height=\maxheight,keepaspectratio}
\makeatletter
\def\fps@figure{htbp}
\makeatother
\providecommand{\tightlist}{%
  \setlength{\itemsep}{0pt}\setlength{\parskip}{0pt}}
\usepackage{fancyhdr}
\usepackage{lastpage}
\ifLuaTeX
  \usepackage{selnolig}  
\fi
\usepackage[]{biblatex}
\IfFileExists{bookmark.sty}{\usepackage{bookmark}}{\usepackage{hyperref}}
\IfFileExists{xurl.sty}{\usepackage{xurl}}{} 
\hypersetup{
  pdftitle={HiQ - A Declarative, Non-intrusive, Dynamic and Transparent Observability and Optimization System},
  pdfauthor={Fuheng Wu; Ivan Davchev; Jun Qian},
  colorlinks=true,
  linkcolor={Maroon},
  filecolor={Maroon},
  citecolor={Blue},
  urlcolor={blue},
  pdfcreator={LaTeX via pandoc}}

\title{HiQ - A Declarative, Non-intrusive, Dynamic and Transparent Observability and Optimization System\thanks{HiQ is developed at OCI vision services team to productionize deep learning models. Corresponding author: \href{mailto:fuheng.wu@oracle.com}{\nolinkurl{fuheng.wu@oracle.com}}. Mailing address: 100 Oracle Pkwy, Redwood City, CA 94065, USA.}}
\usepackage{etoolbox}
\makeatletter
\providecommand{\subtitle}[1]{
  \apptocmd{\@title}{\par {\large #1 \par}}{}{}
}
\makeatother
\subtitle{Designed for Observaberable and Performant Software Systems in Modern AI Era}
\author{Fuheng Wu\footnote{Vision Services, OCI Cloud AI, 100 Oracle Pkwy, Redwood City, CA 94065, USA \href{mailto:fuheng.wu@oracle.com}{\nolinkurl{fuheng.wu@oracle.com}}} \and Ivan Davchev\footnote{Vision Services, OCI Cloud AI, 2300 Oracle Way, Austin, TX 78741, USA \href{mailto:ivan.davchev@oracle.com}{\nolinkurl{ivan.davchev@oracle.com}}} \and Jun Qian\footnote{Vision Services, OCI Cloud AI, 1501 4th Ave \#1800, Seattle, WA 98101, USA, \href{mailto:jun.q.qian@oracle.com}{\nolinkurl{jun.q.qian@oracle.com}}}}
\date{September 23, 2021}

\begin{document}
\maketitle
\begin{abstract}
This paper proposes a non-intrusive, declarative, dynamic and transparent system called \texttt{HiQ} to track Python program runtime information without compromising on the run-time system performance and losing insight. HiQ can be used for monolithic and distributed systems, offline and online applications. HiQ is developed when we optimize our large deep neural network (DNN) models which are written in Python, but it can be generalized to any Python program or distributed system, or even other languages like Java. We have implemented the system and adopted it in our deep learning model life cycle management system to catch the bottleneck while keeping our production code clean and highly performant. The implementation is open-sourced at: \url{https://github.com/oracle/hiq}.
\end{abstract}

\fancypagestyle{plain}{%
  \renewcommand{\headrulewidth}{0pt}%
  \fancyhf{}%
  \fancyhfoffset[L]{1cm} 
  \fancyhfoffset[R]{1cm} 
  \lhead{\bfseries Machine Learning Engineering Paper 2021}
  \fancyfoot[R]{\footnotesize Page \thepage\, of\, \pageref*{LastPage}}
  \setlength\footskip{0pt}
}
\pagestyle{plain}

\hypertarget{introduction}{%
\section{\texorpdfstring{\textbf{Introduction}}{Introduction}}\label{introduction}}

Identifying program performance bottlenecks is crucial for both monolithic applications and distributed systems. Many of these applications and systems are written in Python, a popular programming language famous for its simple syntax and powerful libraries. In many cases, for a given python program or system, tracking the runtime information like latency, memory, I/O is needed, but it is not allowed to modify the original code due to various constraints. In other cases, after logging and tracing code is inserted into the original code, it will inevitably cause performance degradation. How to measure the degradation is a problem. Furthermore, once the code is written and running, another problem is, how can we tune the tracing logic on the fly to get the best system performance without losing insight of the system.

In this paper, we propose a non-intrusive, declarative, dynamic and transparent system called \texttt{HiQ} to track Python program runtime information without compromising on the run-time system performance and losing insight. HiQ is designed for both monolithic applications and distributed systems. We invented this when optimizing our large deep learning models which are written in Python, but it can be generalized to any Python program or distributed system. We have implemented the system and adopted it in our deep learning model life cycle management software \texttt{Gamma} to catch the bottleneck while keeping our production code clean and highly performant.

\hypertarget{background}{%
\section{Background}\label{background}}

\href{https://en.wikipedia.org/wiki/Application_performance_management}{Application performance monitoring/management} has been a common problem for years and has spawned a profitable business for many companies. There are many APM vendors in the market today, such as \href{https://www.oracle.com/manageability/enterprise-manager/technologies/application-performance-management.html}{Oracle APM}, Elastic APM, IBM Instana APM, Splunk APM, to name a few.

For offline performance tracking, people have used tools like Python's built-in library \texttt{cProfile}, \texttt{profile}, \texttt{pstats} for a long time. They are powerful analytics tools but have many drawbacks, including timing accuracy, high overhead, overwhelming irrelevant information, difficult to customize and only useful for development purpose \autocite{scalene}.

For online performance tracking, since Google published its Dapper paper in 2010\autocite{dapper}, there have emerged many different implementations like Zipkin(2012), Jaeger(2016) and Apache SkyWalking(2018). Basically, to tackle the online performance tracking problem, the industry has two methods. The first traditional method, which is also called \texttt{explicit\ instrumentation}, requires developers to manually insert tracing code snippets into the original source code. The second way is to use an agent thread to instrument the original code. When the program starts, in addition to the main thread, an agent thread is spawned to periodically trigger a thread dump and send the thread snapshot to the remote analytics server. The analytics server stores the information, consolidates them and reconstructs the runtime performance information for users to query\autocite{oapm}.

\begin{figure}
\centering
\includegraphics{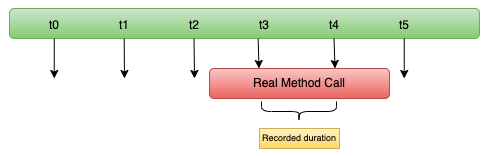}
\caption{Monitoring Blind Spot with Agent\label{skbs}}
\end{figure}

Most of the APM vendors provide the first method as the solution for Python programs. Some of them like Elastic APM also provide libraries with the second method. However, even for the second method, although it doesn't change the original code, it requires complex setup, at least one more running thread and extra network I/O due to the frequent thread dump and communication between the client and data collection server. As for Python programs, especially for CPU bound applications like our case, the overhead is huge because of the design of the Python language. Also reconstructing the runtime information from periodically generated thread snapshots is not as precise as the result from the method one. For instance, in Apache SkyWalking, events with duration less than 10ms will be missed by the system(Figure \ref{skbs}) \autocite{swbs}.

This invention disclosure will propose a new way for non-intrusive performance tracking without extra thread, minimum network I/O overhead but also reach the same accuracy as method one. Other than that, it also proposes a method to control the monitoring overhead dynamically, and a method to measure the tracing overhead and display it to the users.

Existing solutions mostly employ an intrusive \texttt{explicit\ instrumentation} way described above. The existing state-of-the-art solutions like Apache SkyWalking adopt agent-based methods, but with a lot of drawbacks.

Our proposed method solves the same problem with a totally different approach. Our first-of-it-kind solution fully leverages the dynamic features of Python interpreter. The driver code first loads all the variables, functions in interest from external storage, either a configuration file or a database, into memory, and then it loads the target Python code, either script or module. When it loads the target code, \texttt{HiQ} will record the original memory address of the variable or function in interest. Then it will dynamically construct new functions to intercept the invocation to the original function or variables. Several Python decorators are provided so that latency, memory, I/O can be easily traced out of the box. We also provide interfaces for users to define their own tracing functions. When the functions are invoked, a call graph is built in the memory and data is collected and stored in the graph nodes. Another controller process sets all the switches in the shared memory which is accessible from the \texttt{HiQ} driver code, so the behavior of the tracing can be controlled by a remote scheduler. In a distributed environment, the scheduler can schedule the tracing to different levels, like enabling aggressive tracing or disabling part or all of them, for different machines, or adaptively enable them according to the load of machine, or other factors.

We have implemented our algorithm in a Python library called \texttt{HiQ}. It brings the application performance observability to a new level without any compromise of the code completeness, cleanness and run-time performance. It makes our tracking easy, dynamic, non-intrusive, transparent and highly performant.

\hypertarget{hiq-methodology}{%
\section{HiQ Methodology}\label{hiq-methodology}}

\hypertarget{overview}{%
\subsection{Overview}\label{overview}}

HiQ is a \texttt{declarative}, \texttt{non-intrusive}, \texttt{dynamic} and \texttt{transparent} tracking system for both \textbf{monolithic} application and \textbf{distributed} system. It brings the runtime information tracking to a new level which you can never imagine before. And HiQ doesn't compromise with speed and system performance, or hide any tracking overhead information. HiQ applies for both I/O bound and CPU bound applications. It is fully customizable, fully dynamic, and transparent. It could be a development tool, but it is designed with production level applications support in mind, so \textbf{you can use HiQ in both development and production environment}. Also it is flexible, user-friendly and easy to integrate with existing open-source projects like Zipkin, Jaeger, Kafka, Service Mesh, and enterprise-level services like Oracle APM, Oracle Functions, Oracle T2 etc.

\hypertarget{proposed-data-structure-and-system-architecture}{%
\subsection{Proposed Data Structure and System Architecture}\label{proposed-data-structure-and-system-architecture}}

\hypertarget{driver-code-and-target-code}{%
\subsubsection{Driver Code and Target Code}\label{driver-code-and-target-code}}

\begin{figure}
\centering
\includegraphics{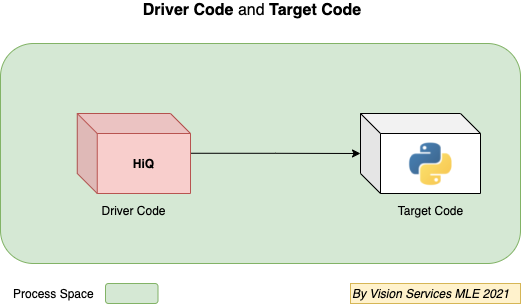}
\caption{A Basic Setup for HiQ Tracing\label{driver}}
\end{figure}

One of the most basic concepts in HiQ are \textbf{driver code} and \textbf{target code}. A typical set up could be illustrated by figure \ref{driver}.

The target code is the original python script or module, which contains the business logic and is importable. The driver code contains the tracking logic. It imports the target code and invokes the function in it.

\hypertarget{declaration-with-minimum-coding}{%
\subsubsection{Declaration With Minimum Coding}\label{declaration-with-minimum-coding}}

Instead of diving into the source code to insert logging and tracing code, HiQ uses a declarative configuration file to describe what you want to trace. It could be a json file with format like:

\begin{Shaded}
\begin{Highlighting}[]
\OtherTok{[}
  \FunctionTok{\{}
    \DataTypeTok{"name"}\FunctionTok{:} \StringTok{"f1"}\FunctionTok{,}
    \DataTypeTok{"module"}\FunctionTok{:} \StringTok{"my\_model\_1"}\FunctionTok{,}
    \DataTypeTok{"function"}\FunctionTok{:} \StringTok{"func1"}\FunctionTok{,}
    \DataTypeTok{"class"}\FunctionTok{:} \StringTok{""}
  \FunctionTok{\}}\OtherTok{,}
  \FunctionTok{\{}
    \DataTypeTok{"name"}\FunctionTok{:} \StringTok{"f2"}\FunctionTok{,}
    \DataTypeTok{"module"}\FunctionTok{:} \StringTok{"my\_model\_2"}\FunctionTok{,}
    \DataTypeTok{"function"}\FunctionTok{:} \StringTok{"func2"}\FunctionTok{,}
    \DataTypeTok{"class"}\FunctionTok{:} \StringTok{""}
  \FunctionTok{\}}\OtherTok{,}
\OtherTok{]}
\end{Highlighting}
\end{Shaded}

The json file contains a list of dictionaries specifying the module, class and function in interest, and the \texttt{name} is the name of the HiQ tree node. The above json can be translated into the instructions like: trace the function \texttt{func1} in module \texttt{my\_model\_1} as name \texttt{f1}, and trace the function \texttt{func2} in module \texttt{my\_model\_2} as name \texttt{f2}.

The user just needs to declare the tracking target and choose one HiQ class, for instance \texttt{HiQLatency} for latency tracking, \texttt{HiQMemory} for memory tracking, then the HiQ system will do the tracking automatically.

\hypertarget{hiq-tree}{%
\subsubsection{HiQ Tree}\label{hiq-tree}}

\begin{figure}
\centering
\includegraphics{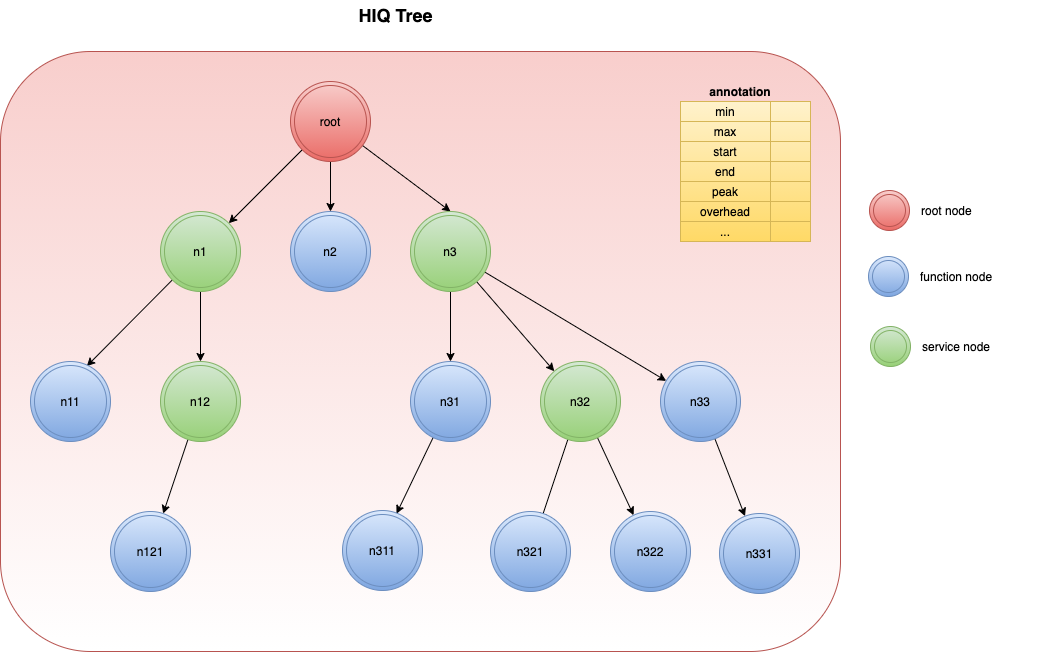}
\caption{HiQ Tree\label{hiqtree}}
\end{figure}

HiQ tree is a special \href{https://en.wikipedia.org/wiki/Interval_tree}{Interval Tree} data structure designed for performance tracking and root-cause finding. For monolithic applications, it is similar to the so-called \texttt{call\ graph}\autocite{callgraph}. For distributed systems, it is similar to \texttt{trace\ tree} which was described in Google's Dapper paper\autocite{dapper}. HiQ tree is for both monolithic and distributed tracing, so it combines features of both \texttt{call\ graph} and \texttt{trace\ tree}. Figure \ref{hiqtree} illustrates how an HiQ looks like.

HiQ tree's features include:

\begin{itemize}
\item
  HiQ tree is composed of two types of subtrees: service subtrees and function subtrees. Service subtree is composed of service nodes, and function subtree is composed of function nodes.
\item
  HiQ tree is configurable, dynamic and can be adaptively tuned according to user requirements like the system performance SLA.

  \begin{itemize}
  \tightlist
  \item
    A configuration file loaded in shared memory will control which process and node should be traced or ignored and in what conditions.
  \item
    The tree structure can dynamically change according to run-time performance information and the configuration. For instance, you can set a HiQ tree to ignore all function nodes with span less than 100 milliseconds. Actually when the function tree is set in \texttt{concise} mode, it will ignore all zero-span nodes. This is very useful to memory tracing because memory cost is not monotonic due to the operating system's memory management operation or program's garbage collection.
  \item
    The tree structure can adaptively change to meet system performance SLA.
  \end{itemize}
\item
  HiQ tree's subtrees are constructed in different ways according to the subtree type. HiQ service subtree is sent to HiQ server and the tree is re-constructed afterwards. HiQ function subtree is built in memory at runtime when the related function calls are finished. The overhead is mitigated with multi-process design and transparent to users. This design also simplifies the tree reconstruction by removing the need of complex reconstructing algorithm\autocite{stam}.
\item
  HiQ tree is built or re-constructed in a non-intrusive way so that business logic and tracing logic are separated. This brings a lot of benefits and flexibility. For instance, we can choose to use different logging libraries at runtime without touching the target code.
\end{itemize}

\hypertarget{tracking-with-transparent-overhead}{%
\subsubsection{Tracking with Transparent Overhead}\label{tracking-with-transparent-overhead}}

Any tracing or logging has overhead. Sometimes, the overhead can be huge, taking Python's \texttt{profile} module as an example. It also depends on how you write your code. Since we are talking about Python, we are only interested in latency overhead. The CPU and memory overhead is either negligible or can be represented by latency. A lot of software declares they have minimum overhead but they don't tell you how minimum it is. It is very likely that you write low performance code with huge overhead but are not aware of it. \texttt{HiQ} is the \textbf{one and only one system to make this completely transparent to users}. HiQ keeps track of latency for the original function and the tracking code. The tracking code runs before and after the original function call, so HiQ is able to calculate the logging code overhead easily. It attaches latency overhead with each HiQ tree, which gives you an understanding of how big or small the impact of the tracking system itself is. The latency overhead can be printed out in absolute format, with a unit of microsecond, or percentage format.

\begin{figure}
\centering
\includegraphics{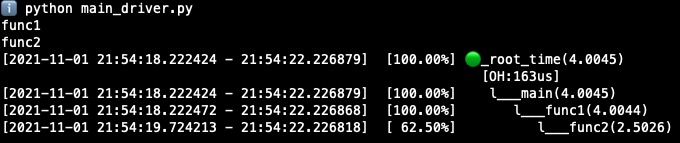}
\caption{Latency Overhead in An HiQ Tree\label{oh}}
\end{figure}

Figure \ref{oh} is an example of how the overhead is attached in the HiQ tree. The tag \texttt{OH:} under the root node displays the latency overhead of the entire tree. We can see the \texttt{main} function took 4.0045 seconds to finish, and the tracing overhead is 163us, or 0.004\%.

\hypertarget{hiq-system-architecture-for-monolithic-application}{%
\subsubsection{HiQ System Architecture for Monolithic Application}\label{hiq-system-architecture-for-monolithic-application}}

\begin{figure}
\centering
\includegraphics{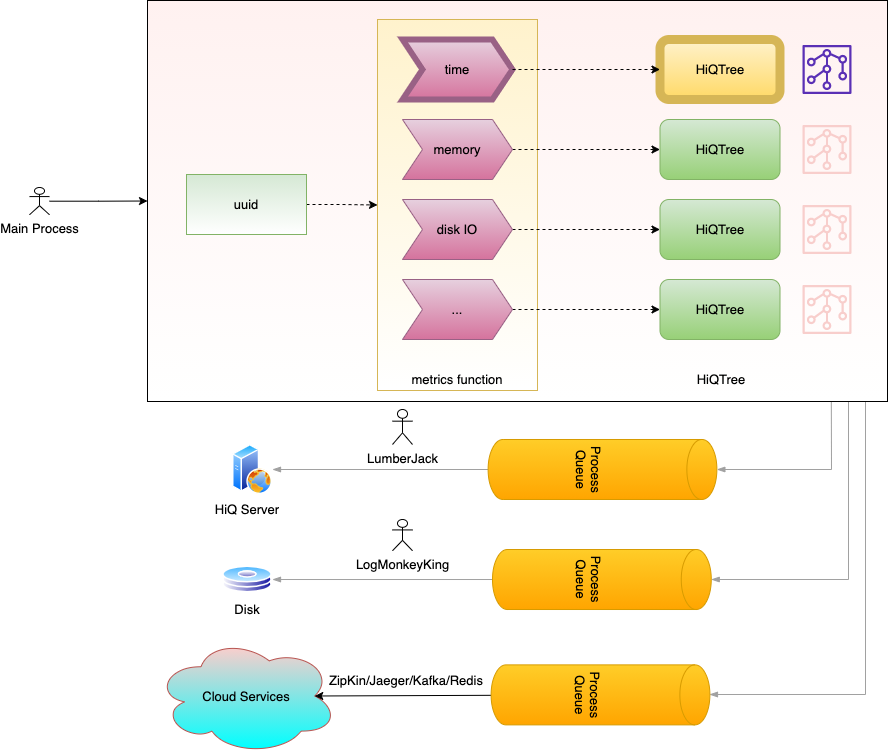}
\caption{HiQ In Monolithic Application\label{hiqmonsys}}
\end{figure}

For monolithic applications, HiQ driver code launches the target code in the main thread of the main process. At the same time, depending on your setup, it also spawns multiple processes for saving logs locally, or sending metrics to HiQ server or cloud services like Oracle APM or Kafka. The system architecture is illustrated at figure \ref{hiqmonsys}.

In the main process, a map data structure is used to host different HiQ Trees for different metrics, including latency, memory, disk I/O. The map has a maximum size. When the map size reaches the maximum value, the HiQ trees will be sent to other processes, via Python multiprocessing queue, for further processing and the map will be reset to empty.

\hypertarget{hiq-system-architecture-for-distributed-system}{%
\subsubsection{HiQ System Architecture for Distributed System}\label{hiq-system-architecture-for-distributed-system}}

\begin{figure}
\centering
\includegraphics{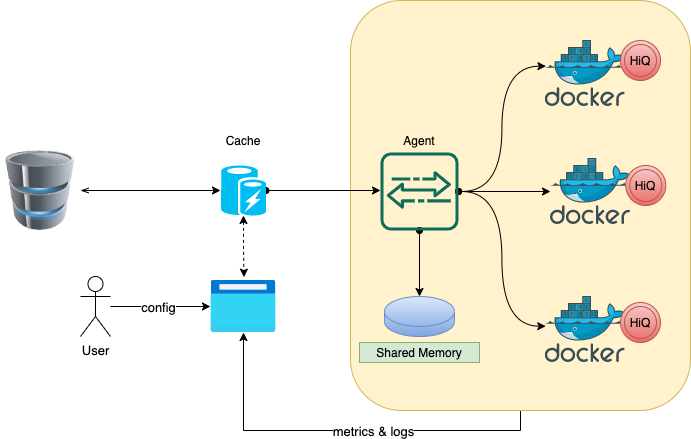}
\caption{HiQ In Distributed System\label{hiqdissys}}
\end{figure}

As for distributed systems, at application level, the HiQ system is the same as that described above. But from a higher distributed system level, HiQ architecture is illustrated by figure \ref{hiqdissys}.

Here we assume each application is running in a docker instance hosted in a VM(virtual machine). This setup is very common in distributed systems managed by Kubernetes or Mesos. For instance, the layout is exactly the same in Kubernetes but they just have different names there: the VM is called \texttt{node}, the docker instance is called \texttt{pod} or \texttt{container}.

So firstly, users set up the configuration for HiQ tracing in a web browser. The data is written into a cache system and then saved persistently into a database. An agent running in a VM will get notified by the cache server for the configuration change if the change is relevant to that VM. Then the agent will pull the change and update the shared memory. The shared memory is used by all the docker instances running in that VM. Once the configuration is changed, the HiQ system will pick up the change accordingly. By this way, the user can control HiQ tracing at run time and at each level: VM, instance, service, and function.

\hypertarget{use-case-examples}{%
\subsection{Use Case Examples}\label{use-case-examples}}

\hypertarget{hiq-for-non-intrusive-monitoring-with-oracle-apm}{%
\subsubsection{HiQ for Non-intrusive Monitoring with Oracle APM}\label{hiq-for-non-intrusive-monitoring-with-oracle-apm}}

\href{https://www.oracle.com/manageability/application-performance-monitoring/}{OCI Application Performance Monitoring (APM)} is a service that provides deep visibility into the performance of applications and enables DevOps professionals to diagnose issues quickly in order to deliver a consistent level of service. \textbf{HiQ supports OCI APM out of the box.}

There are two ways to use OCI APM in HiQ. The legacy way is to use \texttt{HiQOciApmContext} which uses \texttt{py\_zipkin} under the hood. This is the same as what is described in Oracle's official document. The modern way is to use \texttt{HiQOpenTelemetryContext}, which uses the new \texttt{OpenTelemetry} SDK API. The following is a sample code to use HiQ to send metrics to OCI APM, where you don't even need to explicitly write any code for sending metrics, because HiQ has done it automatically for you. What you need to do is to declare the metrics, the module, class and function in interest, and HiQ will do the rest of the heavy lifting.

\begin{figure}
\centering
\includegraphics{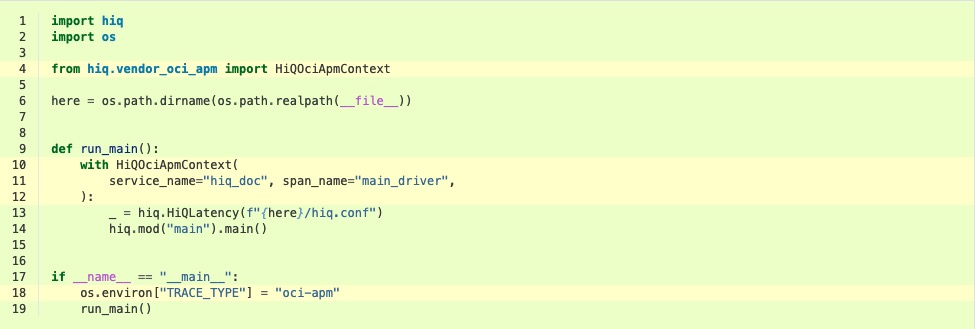}
\caption{HiQ Integration With Oracle APM - Code Sample\label{cs1}}
\end{figure}

\begin{figure}
\centering
\includegraphics{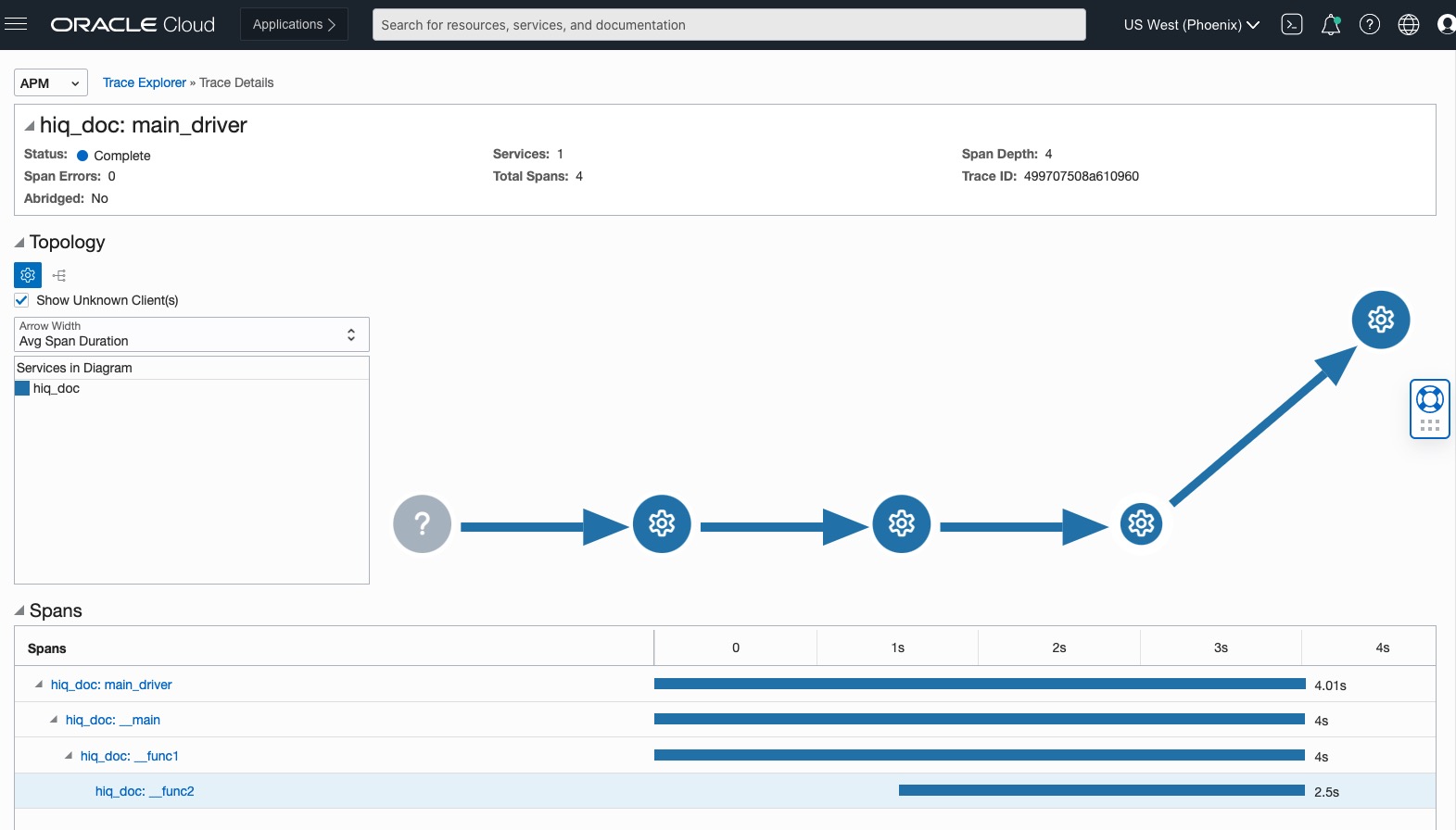}
\caption{HiQ Integration With Oracle APM - Tracing Result\label{apm1}}
\end{figure}

Running this code of figure \ref{cs1} will send metrics to OCI APM and a call graph was displayed as figure \ref{apm1}.

HiQ simplifies the process to send metrics to OCI APM, and it has more benefits than the current method provided by the official Oracle APM documents.

\hypertarget{hiq-in-deep-learning-model-life-cycle-management-system}{%
\subsubsection{HiQ in Deep Learning Model Life Cycle Management System}\label{hiq-in-deep-learning-model-life-cycle-management-system}}

\begin{figure}
\centering
\includegraphics{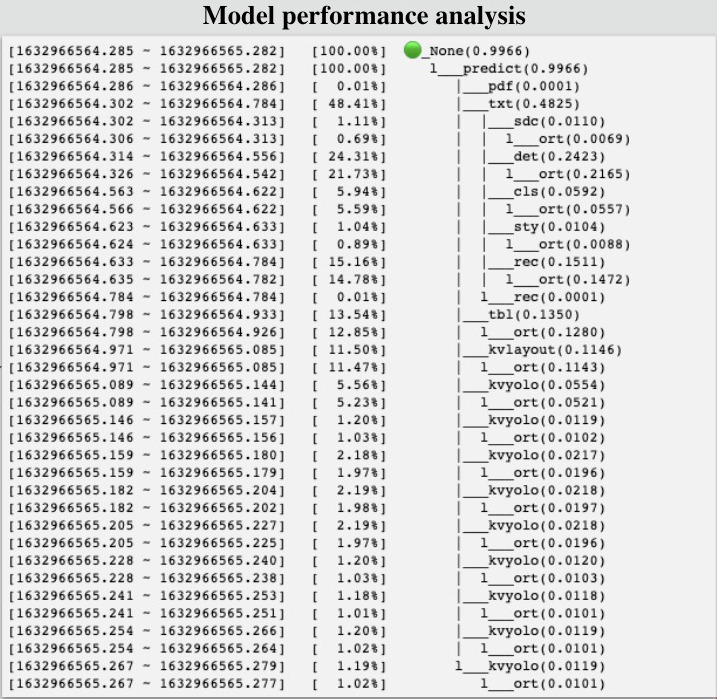}
\caption{HiQ Latency Tree\label{ppt1}}
\end{figure}

HiQ has been widely used in deep learning model life cycle management system to analyze our deep learning model performance. In the analytics web portal, we use HiQ for model latency analysis, which is an implementation of HiQ for monolithic application, where we view a deep learning model as a monolithic software. The figure \ref{ppt1} shows an HiQ latency tree for a sample DNN model. We can see the absolute value and percentage of the time cost for each sub-model. We can clearly see the text detection model took 24\% of the total time and is the bottleneck.

\begin{figure}
\centering
\includegraphics{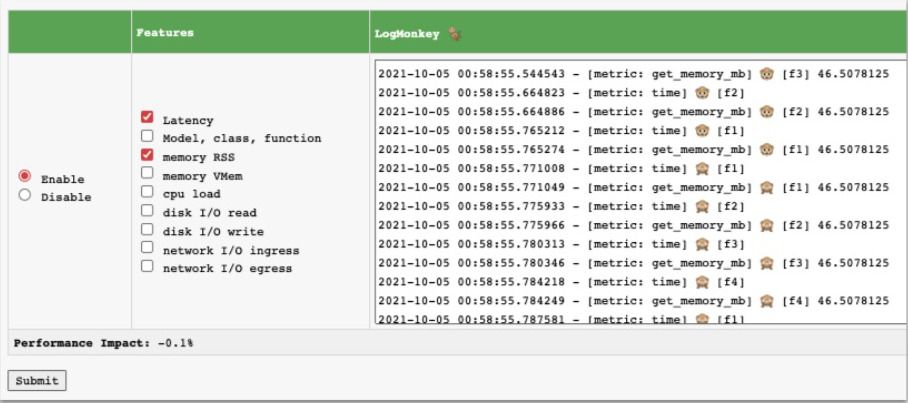}
\caption{Dynamic Setup In Distributed System\label{ppt2}}
\end{figure}

In the analytics platform, we use HiQ in a distributed way, by which the Observability Center can dynamically enable and disable logging for different metrics, modules and functions as figure \ref{ppt2}. The \texttt{Log\ Monkey} console changes run-time log output accordingly. Different from all the products in the market, Observability Center will show the performance overhead incurred by user's selection of logging metrics, so that user can find the best balance between the performance overhead and run-time information.

\hypertarget{hiq-integration-with-jaeger-for-distributed-tracing}{%
\subsubsection{HiQ Integration with Jaeger for Distributed Tracing}\label{hiq-integration-with-jaeger-for-distributed-tracing}}

Jaeger, inspired by Dapper and OpenZipkin, is a distributed tracing platform created by Uber Technologies and donated to Cloud Native Computing Foundation. It can be used for monitoring microservices-based distributed systems. At the time of this writing, Jeager supports two serialization protocols: Thrift and Protobuf. HiQ integrates with Jaeger and both protocols seamlessly.

\begin{figure}
\centering
\includegraphics{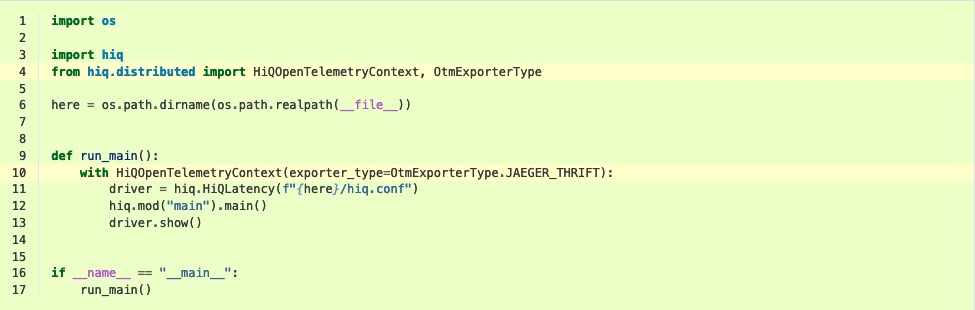}
\caption{HiQ Integration With Jaeger - Code Sample\label{cs2}}
\end{figure}

Figure \ref{cs2} is the driver code for sending metrics data to Jaeger by Thrift and HTTP. You can see, compared with figure \ref{cs1}, the only change is line 4 and 10. You only need to add a context manager \texttt{hiq.distributed.HiQOpenTelemetryContext} to get Jaeger tracing working.

\begin{figure}
\centering
\includegraphics{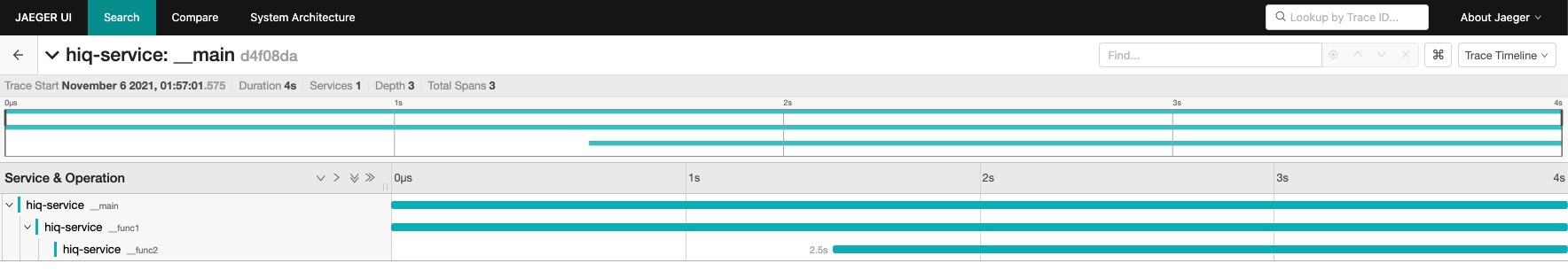}
\caption{HiQ Integration with Jaeger - Tracing Result\label{jaeger1}}
\end{figure}

Run the driver code and check Jaeger UI, we can see the traces have been recorded from figure \ref{jaeger1}.

Other than the use cases above, HiQ supports many frameworks, like ZipKin, Flask, FastAPI, etc out of the box.

\hypertarget{conclusion}{%
\section{Conclusion}\label{conclusion}}

In this paper, we proposed HiQ to provide a declarative, non-intrusive, dynamic and transparent method to track python program performance. It is designed for both monolithic application and distributed system tracing and can be easily integrated into existing software or services. HiQ can be embedded as a new feature of existing APM product and also a brand new way for performance monitoring, root cause finding, and system logging. We have used HiQ in our development, pre-production and production environments and it helped us to find bugs and optimize our services.

\printbibliography

\end{document}